\documentclass[twocolumn, superscriptaddress]{revtex4-2} 
\usepackage[T1]{fontenc}
\usepackage[utf8]{inputenc}
\usepackage{appendix}
\usepackage{amsmath,amssymb}
\usepackage{bm}
\usepackage{graphicx}
\usepackage[dvipsnames]{xcolor}
\usepackage{tikz}
\usepackage[normalem]{ulem}
\usepackage{braket}
\usepackage{siunitx}
\usepackage[version=4]{mhchem}
\usepackage[colorlinks=true, linkcolor=blue, citecolor=blue, urlcolor=blue]{hyperref}
\usepackage[capitalize]{cleveref}
\usepackage{microtype}
\DeclareSIUnit\gauss{G}
\newcommand{\subref}[2][]{\hyperref[#2]{\ref*{#2}(#1)}}

\begin{document}
	
	\title{Quantum statistics on atom-ion Feshbach resonances}
	
	\author{Joachim Siemund}
	\affiliation{Physikalisches Institut, Albert-Ludwigs-Universit\"at Freiburg, Hermann-Herder Str. 3, 79104 Freiburg, Germany}
	\author{Fabian Thielemann}
	\affiliation{Physikalisches Institut, Albert-Ludwigs-Universit\"at Freiburg, Hermann-Herder Str. 3, 79104 Freiburg, Germany}
	\author{Jonathan Grieshaber}
	\affiliation{Physikalisches Institut, Albert-Ludwigs-Universit\"at Freiburg, Hermann-Herder Str. 3, 79104 Freiburg, Germany}
	\author{Wei Wu}
	\affiliation{Physikalisches Institut, Albert-Ludwigs-Universit\"at Freiburg, Hermann-Herder Str. 3, 79104 Freiburg, Germany}
    \author{Patrick Mullan}
	\affiliation{Physikalisches Institut, Albert-Ludwigs-Universit\"at Freiburg, Hermann-Herder Str. 3, 79104 Freiburg, Germany}
	\author{Panagiotis Giannakeas}
	\affiliation{Max Planck Institute for the Physics of Complex Systems, N\"othnitzer Str. 38, 01187 Dresden, Germany}
	\author{Krzysztof Jachymski}
	\affiliation{Faculty of Physics, University of Warsaw, Pasteura 5, 02-093 Warsaw, Poland}
	\author{Tobias Schaetz}
	\affiliation{Physikalisches Institut, Albert-Ludwigs-Universit\"at Freiburg, Hermann-Herder Str. 3, 79104 Freiburg, Germany}
	
	\date{\today}
	
	\begin{abstract}
		We investigate three-body recombination in a hybrid atom-ion system consisting of a single trapped \ce{Ba^+} ion immersed in a two-component Fermi gas of \ce{Li} atoms near an atom-ion Feshbach resonance. By tuning the spin composition at constant density and temperature, we isolate the role of quantum statistics in atom-atom-ion collisions. The measured ion loss rate exhibits a pronounced nonlinear dependence on spin polarization, revealing a reduced contribution of recombination pathways involving identical fermions already at the level of experimental observables. The observations are consistent with a two-step recombination picture and an adiabatic hyperspherical approach, where antisymmetrization restricts the available entrance channels and gives rise to interference between indistinguishable recombination pathways. Our work establishes atom-ion systems as a platform for controlling three-body collisions via quantum statistics and demonstrates that exchange-symmetry effects remain robust even under thermal averaging that obscures the underlying threshold-law behavior.
	\end{abstract}
	
	\maketitle
	\newpage
    
	\section{Introduction}
    Few-body collisions within cold gases provide a sensitive probe of quantum statistics and interaction dynamics~\cite{GiorginiFermiGasTheory2008,greene_review_2017}, where threshold-law physics constrains accessible scattering channels through symmetry-imposed angular-momentum selection rules at low collision energies. In particular, antisymmetrization suppresses $s$-wave contributions for identical fermions, leading to the Pauli blockade of few-body processes. These fundamental constraints have far-reaching implications, from ultracold atomic gases to chemical reactions and their control at low energies.
    In neutral atomic systems, control of two- and three-body collisions allows quantum degeneracy to be reached while maintaining high densities, where threshold laws give rise to characteristic phenomena such as the Efimov effect and Pauli suppression of recombination~\cite{Kraemer2006,Sanner2010,greene_review_2017,naidonEfimovPhysicsReview2017,dincaoFewbodyPhysicsResonantly2018}.
    In multicomponent Fermi gases, three-body recombination has been studied as a function of spin composition~\cite{Huckans2009,ottensteinCollisionalStabilityThreeComponent2008,WenzUniversalTrimer2009,NiemannPauliblockingeffects2012,schumacher2026observation}. The high degree of control over interactions and quantum statistics has enabled quantitative comparisons with theory in well-controlled regimes, establishing a well-understood baseline for few-body physics.
    Hybrid atom-ion systems offer a stringent test of few-body physics going beyond density-averaged measurements ~\cite{LitRydbergPolaron2018,CorrInRydbergTrimers2023, deissColdTrappedMolecular2024,meirDynamicsGroundStateCooled2016}, since a single ion acts as a local probe of the surrounding gas. However, the interplay of long-range interactions and finite collision energies introduces a qualitatively different regime than in the case of neutral ultracold atomic vapors~\cite{grierObservationColdCollisions2009a, zipkesTrappedSingleIon2010, ratschbacherControllingChemicalReactions2012a, dieterleTransportSingleCold2021a, harterSingleIonThreeBody2012,tomzaColdHybridIonatom2019}.
    The long-range $1/r^4$ atom-ion potential prevails over extended length scales, predicted to alter the few-body dynamics~\cite{gebalaUniversalityIonicThreebody2025}, while thermal averaging can obscure the underlying quantum behavior~\cite{tomzaColdHybridIonatom2019}. 
    These challenges are the reason why the quantum regime of cold atom-ion collisions has only recently been experimentally accessible~\cite{feldkerBufferGasCooling2020, weckesserObservationFeshbachResonances2021b}.
    This progress in atom-ion hybrid systems opens a new domain for exploring the role of quantum statistics, where understanding how symmetry constraints manifest is essential for connecting few-body theory with experimentally accessible regimes.
    In this context, the controllable spin composition of the surrounding atomic ensemble provides a way to selectively access distinguishable and indistinguishable collision channels. Yet, it remains unclear to what extent interference between indistinguishable reaction pathways and Pauli-induced suppression persists and remains observable in the presence of long-range interactions and finite collision energies in atom-ion hybrid platforms.
    Establishing such symmetry-protected effects is a strategy towards identifying regimes in which strong atom-ion interactions can be retained while dominant loss pathways are reduced.
    Here, we investigate three-body recombination in an atom-ion system~\cite{weckesserObservationFeshbachResonances2021b} near an atom-ion Feshbach resonance by measuring the ion loss rate in a two-component quantum gas with dependence on its spin composition. We find that the collision temperatures in our system are cold enough such that quantum statistics leads to a pronounced, symmetry-dependent modification of the recombination dynamics.

	\section{Experimental setup}
    
    \begin{figure}[!htbp]
		\centering
		\includegraphics[width=0.93\columnwidth]{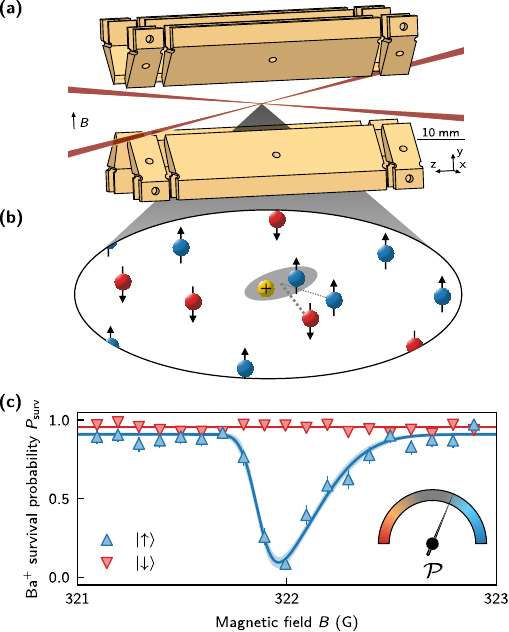}
		\caption{
			\textbf{Polarization-dependent ion loss induced by an atom-ion Feshbach resonance.} (a) Schematic of the atom-ion trap. A single \ce{^{138}Ba+} ion is confined in an rf trap and immersed in ultracold fermionic \ce{^6Li} atoms trapped in a crossed-beam optical dipole trap ($B$ defining the quantization axis). (b) Simplified schematic of the trap center. The \ce{Li} atoms can be prepared in two quasi-spin states, $\ket{\uparrow}$ (blue spheres) and $\ket{\downarrow}$ (red spheres). Near a magnetic Feshbach resonance involving \ce{Ba^+} and $\ce{Li}_{\uparrow}$, resonantly enhanced atom-ion interactions can induce three-body recombination involving a second \ce{Li} atom of either spin state, leading to ion loss (gray area; ion as yellow sphere, spin not depicted; dashed lines indicating recombination pathways).
            (c) \ce{Ba+} survival probability within fully polarized \ce{Li} baths. A pronounced loss signature is observed for $\mathcal{P}=1$ ($\ket{\uparrow}$; blue upward triangles), whereas only weak background loss occurs for $\mathcal{P}=-1$ ($\ket{\downarrow}$; red downward triangles), demonstrating the spin-selective character of the resonance ($t_\mathrm{int}=\SI{200}{\milli\second}$ and $N_\mathrm{tot}\approx 3.8\times 10^4$). The error bars depict $1\sigma$ Wilson score confidence intervals. The solid curves serve as a guide to the eye; the shaded areas denote the $1\sigma$ fit uncertainties. (Inset) Control of bath polarization $\mathcal{P}$ via spin-selective atom-removal.
		}
		\label{fig:fig1}
	\end{figure}
	
	We embed a single Doppler-cooled atomic \ce{^{138}Ba+} ion, of zero nuclear spin, in a cloud of neutral fermionic \ce{^{6}Li} atoms~\cite{schmidtMassselectiveRemovalIons2020,weckesserTrappingShapingIsolating2021}. The ion is trapped in a linear radio-frequency (rf) trap~\cite{leibfriedQuantumDynamicsSingle2003, chenBenchmarkingTrappedionQuantum2024, bermudezAssessingProgressTrappedIon2017,thielemannExploringAtomIonFeshbach2025a} (Fig.~\subref[a]{fig:fig1}). The neutral atoms are confined in a crossed-beam optical dipole trap (xODT) with a temperature of $T_{\ce{Li}}=\SI{0.6(1)}{\micro\kelvin}$ derived from time of flight measurements~\cite{blochQuantumSimulationsUltracold2012, sobireyObservationSuperfluidityStrongly2021, trotzkyTimeResolvedObservationControl2008, andersonObservationBoseEinsteinCondensation1995, demarcoOnsetFermiDegeneracy1999}. The \ce{Li} atoms have a hyperfine structure and are prepared in the $2\,^{2}\mathrm{S}_{1/2}\,\ket{F=1/2}$ ground states $\ket{m_J=-1/2,\,m_I=1}=\ket{\uparrow}$ (lowest energy) and $\ket{m_J=-1/2,\,m_I=0}=\ket{\downarrow}$.
    
	At a magnetic field $B\gtrsim\SI{100}{\gauss}$, we individually address each \ce{Li} hyperfine level with a near-resonant laser beam (see inset of Fig.~\subref[c]{fig:fig1}), applied for a duration $\Delta t_\mathrm{deci}$, to remove atoms from state $\ket{j}$ ($j\in\lbrace \uparrow,\,\downarrow\rbrace$) in a controlled manner. The bath polarization $\mathcal{P}=(N_{\uparrow}-N_{\downarrow})/N_\mathrm{tot}$ and the total atom number $N_\mathrm{tot}=N_{\uparrow}+N_{\downarrow}$ are defined via the atom number $N_j$ of atoms in each state $\ket{j}$.
    
	After preparation of the atomic bath, the ion is axially shuttled to the cloud center, where it sympathetically cools to a median kinetic energy below the \ce{Ba^+-Li} $s$-wave limit, $E_s\approx \SI{8.8}{\micro\kelvin}\times k_\mathrm{B}$~\cite{thielemannExploringAtomIonFeshbach2025a} with the Boltzmann constant $k_\mathrm{B}$. After the atom-ion interaction of duration $t_\mathrm{int}$, the \ce{Li} cloud is released. The ion survival probability, $P_\mathrm{surv}$, is obtained from averaging over several \ce{Ba+} ion fluorescence measurement runs (see Fig.~\subref[c]{fig:fig1}; details in~\cite{weckesserObservationFeshbachResonances2021b, thielemannExploringAtomIonFeshbach2025a}).
    
    \section{Density profile of the two-component Fermi gas}
    
\begin{figure}[t]
        \centering
        \includegraphics[width=\columnwidth]{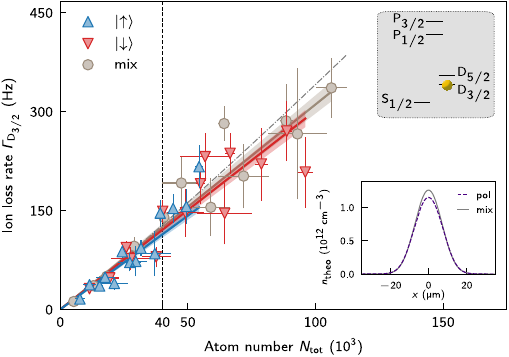}
        \caption{
			\textbf{Local density calibration at the ion position for different spin compositions.} A \ce{Ba^+} ion, optically pumped into the metastable $\mathrm{D}_{3/2}$-level (top-right), undergoes an inelastic two-body collision with one of the surrounding \ce{Li} atoms at the resonance center $B_0$. We measure the loss rate $\Gamma_{\mathrm{D}_{3/2}}$ dependence on $N_\mathrm{tot}$ for three different spin polarizations, ${\mathcal{P}}\in \{-1.0(0);\,-0.24(3);\,1.0(0)\}$ (red triangles; gray circles; blue triangles). All data sets exhibit linear scaling $\Gamma_{\mathrm{D}_{3/2}}\propto N_\mathrm{tot}$, confirming that the local density at the ion position is proportional to $N_\mathrm{tot}$ and independent of ${\mathcal{P}}$ within experimental resolution. Error bars denote $1\sigma$ statistical uncertainties; solid lines represent linear fits with shaded $1\sigma$ confidence intervals. The gray dash-dotted line corresponds to the theoretical mixed gas density calculated with the rates of the polarized gas as a baseline, and is applied as a conservative upper bound in the analysis.
			(Inset) Calculated density profiles $n^\mathrm{theo}_\mathrm{pol}$ and $n^\mathrm{theo}_\mathrm{mix}$ for polarized (${\mathcal{P}}=\pm 1$, dashed) and mixed (${\mathcal{P}}=0$, solid) clouds, respectively. For $N_\mathrm{tot}=40(4)\times 10^3$ atoms, the central density $\hat{n}_\mathrm{tot}$ of the mixed cloud exceeds that of the polarized cloud by $\sim\SI{10}{\%}$, which is not resolved experimentally.
			}
        \label{fig:fig2}
    \end{figure}

    To distinguish spin-dependent three-body processes from density-profile effects, we establish a quantitative mapping between $N_\mathrm{tot}$ and the local density at the ion position $n_\mathrm{tot}=n_{\uparrow}+n_{\downarrow}$. Because the ion probes only a small region of the atomic cloud, it experiences a local polarization that varies across the trap. We perform this calibration using a single \ce{^{138}Ba^+} prepared in the metastable $\mathrm{D}_{3/2}$ state, see \cref{fig:fig2}.

    We measure $P_\mathrm{surv}(t)$ to derive the ion loss rate $\Gamma_{\mathrm{D}_{3/2}}$ that is dominated by two-body inelastic Langevin collisions and is therefore expected to scale linearly with the local total density, independent of $\mathcal{P}$ and the collision energy $E_\mathrm{col}$; deviations from this behavior would indicate a discrepancy in mapping $n_\mathrm{tot}\propto N_\mathrm{tot}$~\cite{xingCompetingExcitationQuenching2024a, jogerObservationCollisionsCold2017}. \cref{fig:fig2} shows the extracted $\Gamma_{\mathrm{D}_{3/2}}$ as a function of $N_\mathrm{tot}$ for both polarized and spin-mixed baths. Within experimental resolution, all data sets exhibit a common linear scaling, $\Gamma_{\mathrm{D}_{3/2}}\propto N_\mathrm{tot}$, which confirms $n_\mathrm{tot}\propto N_\mathrm{tot}$. No significant dependence on $\mathcal{P}$ is observed, indicating that residual \ce{Li-Li} interactions and Fermi pressure do not affect $n_\mathrm{tot}$ within experimental resolution. For the experimental parameters $T_{\ce{Li}}/T_{F}\sim 0.5$, where $T_F$ is the Fermi temperature, theory predicts a maximal density difference of $\sim\SI{10}{\%}$ between polarized and mixed clouds, illustrated in the inset of \cref{fig:fig2}. However, in contrast to the linear density scaling of two-body loss probed here, few-body recombination processes exhibit higher-order scaling in $n_\mathrm{tot}$ and could amplify small deviations. The exact ion position is known with limited accuracy, introducing a potential systematic uncertainty in $n_\mathrm{tot}$. To account for this, we assume a cloud-centered ion position, where the density difference between polarized and mixed clouds ($n^\mathrm{theo}_\mathrm{mix}-n^\mathrm{theo}_\mathrm{pol}$) is maximal. This yields a conservative upper bound on the local density $n_\mathrm{tot}$. The corresponding local polarization, $\hat{\mathcal{P}}=(\hat{n}_{\uparrow}-\hat{n}_{\downarrow})/\hat{n}_\mathrm{tot}$, is evaluated for the same assumed ion position. Any remaining systematic error would weaken, rather than enhance, the observed suppression of the identical-fermion channel.

    \section{Spin-resolved ion loss at a Feshbach resonance}
    
    \begin{figure}[t]
        \centering
        \includegraphics[width=\columnwidth]{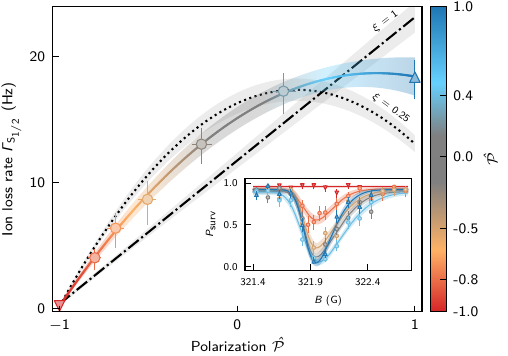}
        \caption{
			\textbf{Polarization dependence of atom-atom-ion recombination at the resonance center.} The ground-state ion loss rate $\Gamma_{\mathrm{S}_{1/2}}$ is shown as a function of the central spin polarization $\hat{\mathcal{P}}$, while keeping the local density at the ion position constant. Error bars denote $1\sigma$ uncertainties from the curve fits. The solid lines show a fit $\Gamma_{\mathrm{S}_{1/2}}(\hat{\mathcal{P}})=a\hat{\mathcal{P}}^2+b\hat{\mathcal{P}}+c$, in agreement with the two-step model. The dash-dotted and dotted lines correspond to the same functional form with fixed ratios of a derived rate equation $\xi=L_{\uparrow\uparrow}/L_{\uparrow\downarrow}=1$ and $\xi=0.25$. The pronounced non-linear dependence on $\hat{\mathcal{P}}$ demonstrates that the loss cannot be described by a single-channel contribution and reflects the relative weight of identical- and distinguishable-fermion three-body recombination pathways. (Inset) $P_\mathrm{surv}(B)$ for different $\hat{\mathcal{P}}$ (indicated by the color bar) at interaction duration $t_\mathrm{int}=\SI{150}{\milli\second}$. Solid lines are skewed Gaussian fits with shaded $1\sigma$ uncertainty regions.
            }
        \label{fig:fig3}
    \end{figure}
    
    Having established a spin-independent density calibration, we now investigate the polarization dependence of the atom-atom-ion interactions in the quantum regime, at the center of the $s$-wave Feshbach resonance at $B_0=\SI{321.90(3)}{\gauss}$~\cite{thielemannExploringAtomIonFeshbach2025a}. We prepare an ion in the electronic ground state $\mathrm{S}_{1/2}$ and measure the resulting ion loss rate $\Gamma_{\mathrm{S}_{1/2}}(\hat{\mathcal{P}})$. In contrast to \cref{fig:fig2}, we keep $\hat{n}_\mathrm{tot}$ constant, while we vary $\hat{\mathcal{P}}$ over its full range $[-1;\,1]$ (see \cref{fig:fig3}). Note that the non-resonant background ion loss rate contribution (a few percent of $\Gamma_{\mathrm{S}_{1/2}}(\hat{\mathcal{P}})$) is neglected. For $\hat{\mathcal{P}}=-1$, no significant resonant loss is observed, indicating that recombination requires the presence of at least one $\ce{Li}_{\uparrow}$ atom. As $\hat{\mathcal{P}}$ increases, $\Gamma_{\mathrm{S}_{1/2}}(\hat{\mathcal{P}})$ exhibits a pronounced non-linear dependence: the loss rate increases towards balanced mixtures ($\hat{\mathcal{P}}\sim 0$), and saturates for $\hat{\mathcal{P}}>0$, where the resonant spin component dominates. The observed nonlinearity directly implies that the measured loss cannot be explained by a purely two-body mechanism or by the density of a single spin component. To quantify this behavior, we first adopt a minimal phenomenological description without assigning microscopic interpretations. We fit the data with the lowest-order polynomial that captures the observed curvature,
    
    \begin{equation}
        \Gamma_{\mathrm{S}_{1/2}}(\hat{\mathcal{P}})=a\hat{\mathcal{P}}^2 + b \hat{\mathcal{P}} +c.
    \end{equation}
    
    A cubic term is not statistically significant and therefore omitted. The fit yields $a=\SI{-6(1)}{\Hz}$, $b=\SI{9.1(8)}{\Hz}$, and $c=\SI{15.1(9)}{\Hz}$. Notably, both linear and quadratic contributions are statistically significant and of comparable magnitude, reflecting the sensitivity of the loss process to the spin composition of the bath. While the fully polarized case $\hat{\mathcal{P}}=1$ isolates the recombination involving identical fermions, the balanced mixture $\hat{\mathcal{P}}=0$ contains additionally a contribution from the process involving two different \ce{Li} states. As a result, the polynomial coefficients cannot be interpreted as direct measures of individual channels. Instead, they encode specific linear combinations of the underlying processes, necessitating an explicit mapping to extract physical rate constants. For three-body processes, this functional form motivates a rate-equation description: at fixed $\hat{n}_\mathrm{tot}$, the loss can be described by a general rate equation of the form 
    
    \begin{equation}
        \Gamma_{\mathrm{S}_{1/2}}=L_{\uparrow\uparrow}\hat{n}_{\uparrow}^2+L_{\uparrow\downarrow}\hat{n}_{\uparrow}\hat{n}_{\downarrow}.
    \end{equation}
    
    Here, $L_{\uparrow\uparrow}$ and $L_{\uparrow\downarrow}$ denote effective three-body recombination rate constants for the $FFX$ ($\ce{Li}_{\uparrow}\ce{Li}_{\uparrow}\ce{Ba+}$) and $FF^\prime X$ ($\ce{Li}_{\uparrow}\ce{Li}_{\downarrow}\ce{Ba+}$) entrance channels, respectively. Naturally, this leads to a quadratic form in $\hat{\mathcal{P}}$, once the densities are expressed in terms of $\hat{n}_{\uparrow/\downarrow}=\hat{n}_\mathrm{tot}(1\pm\hat{\mathcal{P}})/2$. This reveals that the coefficients $a,b,c\propto \hat{n}_\mathrm{tot}^2$ are linear combinations of $L_{\uparrow\uparrow}$ and $L_{\uparrow\downarrow}$, yielding a density-independent ratio $L_{\uparrow\uparrow}/L_{\uparrow\downarrow}=1/(1-2a/b)=0.44(5)$. This implies a reduced relative contribution of the $FFX$ process. To connect this result to a microscopic mechanism, we first consider a Feshbach-assisted two-step recombination process, established for neutral atoms~\cite{hirzlerTrapAssistedComplexesCold2023a,robertsResonanceTheoryTermolecular1969,orelNascentVibrationalRotational1987}, and adapt it to the atom-ion system (see \hyperref[EndMatter]{End Matter}). In this picture, the ion first forms a weakly bound dimer with a ${\ce{Li}}_{\uparrow}$ atom at a rate $\Gamma_\mathrm{dimer}$. This intermediate state subsequently relaxes via a collision with a second \ce{Li} atom of either spin. The corresponding dimer relaxation rate constants $L_{\uparrow}$ and $L_{\downarrow}$ determine the relative weight of the two recombination pathways and thus the observed ion loss rate. Neglecting the possible energy dependence of these parameters, a fit to the experimental data yields
    
    \begin{equation}
        L_{\uparrow}/L_{\downarrow}=0.53^{+0.07}_{-0.08},
    \end{equation}
    
    in agreement with the phenomenological ratio $L_{\uparrow\uparrow}/L_{\uparrow\downarrow}=0.44(5)$. In this framework, both descriptions depend on the relative probability that the second collision occurs with either spin component, allowing a direct mapping between the effective rate constants, relating $L_{\uparrow\uparrow}\mapsto L_{\uparrow}$ and $L_{\uparrow\downarrow}\mapsto L_{\downarrow}$. This consistency shows that the observed suppression of the identical channel reflects the underlying recombination dynamics of an intrinsic three-body process, rather than an artifact of the phenomenological description. At the same time, it emphasizes the role of the second atom in the recombination dynamics: although atom-atom interactions remain negligible in the isolated Fermi gas (see \cref{fig:fig2}), the quantum statistics of the gas become decisive in the presence of the ion, governing the relative contributions of the recombination channels. The experimentally extracted ratio thus provides direct insight into the underlying microscopic dynamics.

    \section{Hyperspherical treatment of TBR}

    \begin{figure}[t]
        \centering
        \includegraphics[width=\columnwidth]{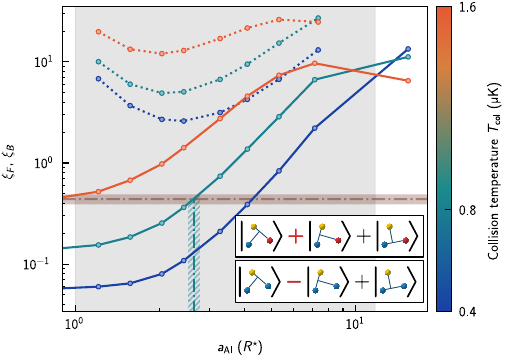}
        \caption{
			\textbf{Experimental-theoretical comparison of the symmetry-dependent three-body loss ratios.} Thermally averaged loss rate constant ratios $\xi_F= L^{FFX}_{3}/L^{FF^\prime X}_{3}$ are shown as functions of the atom-ion scattering length $\text{a}_\mathrm{AI}$ for several center-of-mass collision temperatures $T_\mathrm{col}$ (solid lines, color coded). The experimental temperature is $T_\mathrm{col}=\SI{0.8(1)}{\micro\kelvin}$. The shaded horizontal band indicates the experimentally extracted value of $\xi$ with its $1\sigma$ uncertainty. The intersection determines $\text{a}_\mathrm{AI}\approx 2.7(1)\,R^\star$. The gray region marks the regime $R^\star <\text{a}_\mathrm{AI}<\lambda_\mathrm{th}$, bounded by the atom-ion length scale $R^\star$ and the thermal de Broglie wavelength $\lambda_\mathrm{th}$. Dotted lines show the corresponding bosonic prediction $\xi_{B}= L^{BBX}_{3}/L^{BB^\prime X}_{3}>1$, where $BBX$ and $BB^\prime X$ denote the bosonic counterparts of the $FFX$ and $FF^\prime X$ channels. 
            The comparison shows that the observed suppression arises from fermionic exchange symmetry after thermal averaging.
            (Inset) Simplified Jacobi representation of recombination pathways illustrating the modification of contributing channels for identical fermions compared to distinguishable particles.
		}
        \label{fig:fig4}
    \end{figure}

    Although the two-step picture provides a generic and intuitive description of the data by incorporating magnetic field-dependent resonant enhancement, it also indicates the limits of a model in which the third particle enters only as a secondary collision partner. A framework that treats all three particles at once can provide deeper understanding and potentially requires fewer fit parameters. To this end, we solve the three-body scattering problem in the center of mass frame using hyperspherical coordinates (see \hyperref[EndMatter]{End Matter}), incorporating exchange-symmetry constraints and the associated centrifugal barriers, as well as the dependence on the unknown atom-ion scattering length. \cref{fig:fig4} shows the resulting ratio of the thermally averaged loss rate constants for identical and distinguishable fermionic channels, defined as $\xi_F= L^{FFX}_{3}/L^{FF^\prime X}_{3}$, across a range of experimentally realistic parameters. For identical fermions, exchange symmetry requires antisymmetry of the spatial wave function. This excludes the lowest three-body total angular momentum ($\Lambda=0$) for the relative motion of atoms and enforces odd partial waves among them, yielding $\Lambda=1$ as the leading contribution. This results in a centrifugal-like barrier in the three-body sector and a corresponding node in the three-body wave function, which in the low energy limit yields a suppressed three-body probability density at short-range. This means that at fixed collision energy, the microscopic recombination rate constant $K^{FFX}_{3}$ is smaller than $K^{FF^\prime X}_{3}$, and in the zero-energy limit $K^{FFX}_{3}$ vanishes completely. However, under experimental conditions there is no singular collisional energy, thus the recombination rate constants must be thermally averaged over a Maxwell-Boltzmann collision energy distribution. This smears the barrier-induced suppression, causing the corresponding rate constants $\langle K^{FFX}_{3} \rangle_T$ and $\langle K^{FF^\prime X}_{3} \rangle_T$ to be comparable.
    To connect with the experimentally measured loss rate constants, we account for fermionic indistinguishability, yielding $L^{FFX}_{3} = \frac{1}{2}\langle K^{FFX}_{3} \rangle_T$ and $L^{FF^\prime X}_{3} = \langle K^{FF^\prime X}_{3} \rangle_T$. Microscopically, the factor $1/2$ originates from the number of distinct triplets of particles that contribute to the $FFX$ process. The same factor also occurs for the bosonic $BBX$ process (see \cref{fig:fig4}). In essence, this prefactor ensures that in the rate equations, we do not double-count recombination events with identical particles. Note that this factor has not been imposed a priori in the phenomenological analysis (see \cref{fig:fig3}), but follows directly from the microscopic rate equation and yields quantitative agreement with the experimentally extracted ratio.
    We evaluated four configurations, $FFX$, $FF^\prime X$, $BBX$, and $BB^\prime X$ (with $FF^\prime X$ and $BB^\prime X$ being equivalent). For identical bosons, exchange symmetry leads to constructive interference and enhancement of $\xi_B= L^{BBX}_{3}/L^{BB^\prime X}_{3}>1$. This comparison therefore identifies fermionic symmetry as the dominant mechanism of the observed loss ratio. The mapping further yields an effective atom-ion scattering length $\text{a}_\mathrm{AI}\approx 2.7(1)\,R^\star$, where $R^\star$ is the characteristic atom-ion length scale.
    This value lies in a physically realistic regime, consistent with the predictions of the adiabatic hyperspherical approach. This agreement provides a quantitative experimental constraint on the microscopic three-body description and identifies fermionic exchange symmetry as the dominant mechanism governing the observed loss ratio. More generally, our results show that Pauli-imposed constraints on the accessible Hilbert space can remain experimentally observable even when signatures of centrifugal-barrier suppression are obscured by thermal averaging. This establishes atom-ion systems as a platform for controlling few-body interactions via quantum statistics.
    
\section{Conclusion and Outlook}

We observe that the three-body atom-atom-ion recombination rate is lower for identical fermions than for distinguishable ones. This suppression is robust against the experimentally characterized systematic effects and cannot be explained by threshold-law physics. 
Although conceptually distinct, the phenomenological, two-step, and hyperspherical descriptions are in quantitative agreement with the corresponding measurements, providing an experimental constraint on the microscopic dynamics of atom-ion three-body recombination. Together, they identify a Pauli-induced restriction of the accessible three-body Hilbert space as the dominant mechanism underlying the observed suppression.
The comparison with the hyperspherical calculations indicates that the experiment already operates at the cusp of the threshold-law regime.
With the broad spectrum of atom-ion Feshbach resonances already identified in this system, the dependence of the recombination ratio on the underlying scattering parameters can be explored systematically.
A next step is to explore the crossover from quantum-statistical suppression to the enhancement of identical-fermion recombination predicted by the hyperspherical model. Increasing the collision energy within the current platform introduces additional contributing partial waves, redistributing the quantum scattering weights across the channels while preserving the underlying symmetry constraints. 
This provides a direct experimental benchmark for microscopic three-body descriptions beyond the threshold-law regime and offers the possibility of disentangling the roles of intermediate-state formation and subsequent relaxation.
More generally, the observed suppression identifies regimes in which strong atom-ion interactions persist while dominant loss channels are reduced, offering the prospect of resolving microscopic collision pathways that are currently hidden by thermal averaging.

Looking further ahead, circumventing rf-induced heating via all-optical trapping of ions and atoms promises access to lower and more precisely controlled collision energies~\cite{schmidtOpticalTrapsSympathetic2020,schaetzTrappingIonsAtoms2017}, enabling tunable few-body collisions at the single-particle level. Combined with quantum-statistical control, this may allow engineering recombination processes via St\"uckelberg-type interference between reaction pathways~\cite{Giannakeas2026},
further reducing loss while maintaining strong interactions and ultimately extending atom-ion systems towards controlled impurity physics and charged-polaron realizations.
	
	\section{Acknowledgments}
    This project has received funding from the European Research Council (ERC) under the European Union's Horizon 2020 research and innovation program (Grant No. 648330), the Deutsche Forschungsgemeinschaft (DFG, Grant No. SCHA 973/9-1-3017959 and SCHA 973/10-1), and the Georg H. Endress Foundation. J.~S., F.~T., J.~G., and T.~S. acknowledge financial support from the DFG via the RTG DYNCAM 2717. W.~W. acknowledges financial support from the QUSTEC programme, funded by the European Union's Horizon 2020 research and innovation program under the Marie Sk\l{}odowska-Curie (Grant No. 847471). K.~J. was supported by the Polish National Agency for Academic Exchange (NAWA) via the Polish Returns 2019 program. P.~M. and T.~S. acknowledge financial support from the Georg H. Endress foundation.

    \section{Authorship}
    J.~S. and F.~T., under the supervision of T.~S., jointly developed the experimental methodology, analyzed the data, and interpreted the results. J.~S. carried out the experiments and performed the Fermi-Dirac simulations. J.~G. supported experimental measurements and data acquisition. J.~S., F.~T., J.~G., and W.~W. maintained and further developed the experimental apparatus. K.~J. developed the two-step model and performed the corresponding theoretical calculations. P.~G. developed the hyperspherical approach and carried out the associated theoretical analysis. The manuscript was prepared by J.~S., P.~M., and T.~S., with input from all authors. T.~S. supervised the project. All authors contributed to scientific discussions and to the interpretation of the results. 
	
	\section{Data and code availability}
    The data supporting the findings of this article are not publicly available. The data are available from the authors upon reasonable request.
	
	\bibliography{main}

	\clearpage
	
	\appendix
	
\section{Adiabatic hyperspherical potentials}\label{EndMatter}
The relative degrees of freedom of an arbitrary three-body system can be conveniently described
utilizing the hyperspherical coordinates (for details see~\cite{greene_review_2017}), yielding the following Hamiltonian
\begin{equation}
	\label{eq:eq2}
	H=-\frac{\hbar^2}{2 \mu R^{5/2}} \frac{\partial^2}{\partial R^2}[R^{5/2}\cdot~~] + \frac{\hbar^2}{2 \mu} \hat{\Lambda}^2 + \frac{15\hbar^2}{8\mu R^2} +\sum_{i>j}V_{ij}(\Omega;R),
\end{equation}
where $\mu=m_A\sqrt{ m_I/(2m_A+m_I)}$ is the three-body reduced mass, $m_A$ ($m_I$) refers to the mass of the atom (ion), $R$ is the hyperradius which controls the overall size of the three-body system, $\Omega$ indicates a collective coordinate denoting the five hyperangles~\cite{avery_hyperspherical_1989,smirnov_method_1977}, and $\hat{\Lambda}$ denotes the grand angular momentum operator. The atom-atom and ion-atom interactions are modeled by the following expressions~\cite{poschlBemerkungenZurQuantenmechanik1933,higginspra2022}:
\begin{subequations}
	\begin{align}
		&V_{AA}(r_{AA})=-\frac{D}{\cosh^2(r_{AA}/r_0)} \label{aa_pot}\\
		&V_{IA}(r_{IA})=-\frac{C_4}{r_{IA}^4}(1-e^{-\frac{r_{IA}^2}{\alpha^2}})^3, 
		\label{ia_pot}
	\end{align}
\end{subequations}
where $r_{AA}$ ($r_{IA}$) corresponds to the relative atom-atom (ion-atom) coordinate, $r_0$ indicates the range of the atom-atom interaction and is set equal to $R_\mathrm{vdW}=31.235~a_0$ matching the van der Waals length scale~\cite{chinFeshbachResonancesUltracold2010} for two \ce{^6Li},
$C_4=82~\rm{au}$ is the dispersion coefficient of the ion-atom polarization potential~\cite{puchalskiLithiumElectricDipole2011}, and
the parameters $D$ and $\alpha$ are set such that \cref{aa_pot} and \cref{ia_pot} contain at most a single two-body bound state.

In order to tackle \cref{eq:eq2}, the adiabatic hyperspherical representation is employed, which in a Born-Oppenheimer spirit treats the hyperradius as a slow coordinate whereas $\Omega$ represents the fast degrees of freedom. The total three-body wave function is written as
\begin{equation}
	\Psi(R,\Omega)= R^{-5/2}\sum_\nu \phi_\nu(\Omega;R) F_\nu(R),
	\label{eq:eq4}
\end{equation}
where $F_\nu(R)$ and $\phi_\nu(\Omega;R)$ indicate the $\nu$-th hyperradial and hyperangular part of the wave function, respectively.
In particular, $\phi_\nu(\Omega;R)$ depends only parametrically on the hyperradius and is obtained by diagonalizing \cref{eq:eq2} at fixed $R$
\begin{equation}
	H_{\Omega}(R) \phi_\nu(\Omega;R)=U_\nu(R)\phi_\nu(\Omega;R),
	\label{eq:eq5}
\end{equation}
where the eigenvalues $U_\nu(R)$ are adiabatic hyperspherical potential curves obtained for a given symmetry $\Lambda^{\Pi}$ of the three-body system that possesses total angular momentum $\Lambda$ in the body-fixed frame and inversion parity $\Pi$. Specifically, for systems with three distinguishable particles most dominant contributions in collisions are attributed to $0^+$ symmetry, whereas in the case of two identical fermions the $1^-$ symmetry prevails. Finally, substitution of \cref{eq:eq4,eq:eq5} into the three-body Schr\"odinger equation and integration over all hyperangles $\Omega$ yields a set of coupled ordinary second-order differential equations that solely depend on the hyperradius $R$ and contain all the relevant non-adiabatic couplings (for details see~\cite{greene_review_2017}). In the calculations shown in \cref{fig:fig4} of the main text we use the two lowest hyperspherical potential curves for $FF^\prime X$, $FFX$ and $BBX$.

\section{Three-body recombination}

Solving the multichannel scattering problem formulated above enables the calculation of recombination rates. Note that in this case there is no Feshbach resonance present, instead the process is controlled by the binding energy of a shallow dimer state. It is rather well established that for ion-atom systems the dominating loss process is due to production of molecular ions. The recombination rate constant $K_3$ is proportional to the off-diagonal element of the scattering matrix $K_3\propto \frac{\hbar\pi^2}{\mu k^4}|S_{12}|^2$, where $k$ is the wave vector of the incident three-body state in hyperspherical coordinates, and we limit ourselves to the inelastic process leading from the initial state 1 to the final state 2. In the low energy threshold regime, the dynamics is dominated by the long-range behavior of the interactions, which leads to the established Wigner threshold laws for the inelastic rate constants~\cite{dincaoScatteringLengthScaling2005b} 
\begin{align}
	K_3\propto k^{2\Lambda}\left|\text{a}\right|^{2\Lambda+4} \label{eq:K3_proportionality}
\end{align}
with the two-body scattering length $\text{a}=\text{a}_{FX}$ between the \ce{Ba+} ion $X$ and the $\ce{Li}_{\uparrow}$ atom $F$ and $\Lambda$ being the total angular momentum of the effective three-body system. Crucially, for identical fermions, we have $\Lambda=1$ while for distinguishable particles at threshold $\Lambda=0$. This results in a simple scaling law for the ratio between the two processes being 
\begin{align}
	\label{eq:threshold}
	\frac{K_3^{{FF}{X}}}{K_3^{{FF}^\prime {X}}}=\frac{6 k^2\left|\text{a}\right|^{6}}{k^0\left|\text{a}\right|^{4}}=6k^2\text{a}^2
\end{align}
where $K_3^{{FF}{X}}=K_3\left(\Lambda=1\right)$ and $K_3^{{FF}^\prime {X}}=K_3\left(\Lambda=0\right)$. This already hints that at sufficiently low energies, there should be a large difference between the two rate constants. Numerical calculations within the hyperspherical approach accurately reproduce these threshold laws. However, finite energy effects need to be taken into account. Here, we employ thermal energy distributions, which are close to numerical simulations that include realistic experimental conditions.

A complementary approach to describing the three-body loss rate relies on the two-step model, where the scattering occurs through an intermediate resonant state, which is then quenched to the final product by the third particle. This results in the well-known Breit-Wigner profile
\begin{equation}
	|S_{12}|^2_{\uparrow/\downarrow}=\frac{\Gamma_{\uparrow/\downarrow}(E)\Gamma_{\rm dimer}(E)}{(E-E_0)^2+(\Gamma_{\uparrow/\downarrow}(E)+\Gamma_{\rm dimer}(E))^2/4}\, .
\end{equation}
both for the ${FF}{X}$ and ${FF}^\prime {X}$ processes, marked by the spin up or down arrows. Here $E_0$ is the energy of the intermediate resonant dimer, in the case of a magnetic Feshbach resonance given by $\delta\mu(B-B_0)$, $\Gamma_{\rm dimer}$ is the dimer formation rate, and $\Gamma_{\uparrow/\downarrow} = L_{\uparrow/\downarrow} \hat{n}_{\uparrow/\downarrow}$ is its decay rate which depends on the state of the secondary atom. As we are interested in resonant recombination, one may set $B$ to $B_0$. The $\Gamma$ parameters also depend on the collision energy. The formation rate $\Gamma_{\rm dimer}$ should be well described by two-body physics, leading to $\Gamma_{\rm dimer}\propto E^{2\ell+1}$ with $\ell$ being the two-body orbital angular momentum. On the other hand, the breakup rate comes from a three-body process. In one limit it may be treated as a reactive collision between an atom and a loosely bound molecular ion, for which the Langevin rate should provide a good approximation irrespective of the spin state. However, as the resonant dimer is not bound, one might argue that this process proceeds like non-resonant recombination of free particles, obeying the threshold laws mentioned above. Note that within the two-step model the only possibility to enforce the threshold behavior that incorporates fermionic symmetry is to introduce the dependence on the three-body angular momentum $\Lambda$ in $\Gamma_{\uparrow/\downarrow}$. In the last step $|S_{12}|^2_{\uparrow/\downarrow}$ matrix elements can be directly translated into the respective rate constants $K_3^{{FF}{X}}$ and $K_3^{{FF}^\prime {X}}$, and linked to the observable loss rate constants $L_3$.

\end{document}